\documentclass[showpacs,preprintnumbers,amsmath,amssymb]{revtex4}
\usepackage{graphicx}
\usepackage{dcolumn}% Align table columns on decimal point
\usepackage{bm}% bold math

\begin{document}

\title{Generalized Poland-Scheraga model for DNA hybridization}
\author{Thomas Garel
%\footnote{Member of CNRS} 
and Henri Orland \\
%EndAName
Service de Physique Th\'{e}orique, CEA/DSM/SPhT\\
Unit\'e de recherche associ\'ee au CNRS\\
91191 Gif-sur-Yvette cedex, France}
%\maketitle

\begin{abstract}
The Poland-Scheraga (PS) model for the helix-coil transition of DNA
considers the statistical mechanics of the binding (or hybridization) of two
complementary strands of DNA of equal length, with the restriction that only
bases with the same index along the strands are allowed to bind. In this
paper, we extend this model by relaxing these constraints: We propose a
generalization of the PS model which allows for the binding of two strands of
unequal lengths $N_{1}$ and $N_{2}$ with unrelated sequences. We study in
particular (i) the effect of mismatches on the hybridization of
complementary strands (ii) the hybridization of non complementary strands
(as resulting from point mutations) of unequal lengths $N_{1}$ and $N_{2}$.
The use of a Fixman-Freire scheme scales down the computational complexity
of our algorithm from $O(N_{1}^{2}N_{2}^{2})$ to $O(N_{1}N_{2})$.The
simulation of complementary strands of a few kbps yields results almost
identical to the PS model. For short strands of equal or unequal lengths,
the binding displays a strong sensitivity to mutations. This model may be
relevant to the experimental protocol in DNA microarrays, and more generally
to the molecular recognition of DNA fragments. It also provides a
physical implementation of sequence alignments.

\bigskip

%PACS numbers: 87.14.Gg; 87.15.Cc; 82.39.Pj

\end{abstract}
\maketitle
\section{Introduction}

Natural DNA exists as a double helix bound state \cite{CW}. Upon
heating, the two complementary strands may separate. This unbinding
transition is called DNA denaturation (see \cite{Pol_Scher2}). The
reverse process of binding is called renaturation, recombination, or,
in more biological words, recognition between strands. 

An important puzzle is how the extreme selectivity required by the
biological machinery can be achieved in spite of the very high entropy of
non selective binding. For instance, in a DNA microarray, single strands of
DNA are grafted on a surface. When this array is immersed in a solution
containing complementary and mutated strands, the recognition process occurs
with a high accuracy, with a seemingly low rate of errors \cite{chip1,chip2}.

The Poland-Scheraga (PS) model \cite{Pol_Scher1,Pol_Scher2,Poland,Poland2} 
aims at
describing DNA denaturation in a simplified way: In a nutshell, the double
helix is described as a succession of bound fragments separated by unbound
bubbles (loops). More specifically, the PS model assumes that the two DNA
strands are exactly complementary, and that only bases with the same index
can form pairs (implying that the strands have equal lengths $N$). As will
be shown below, its computational time scales like $N^{2}$. Use of the
Fixman-Freire scheme \cite{Fix_Fre,Yer1} reduces the scaling to $O(N)$,
allowing for the study of melting of long sequences (up to a few Mbps)\cite%
{Meltsim}. For a general review on DNA denaturation, we refer the reader to 
\cite{WB85}. 
%In particular, this model can be used to detect coding regions
%in linear (non circular) DNA, and thus gives an independent prediction of
%genes 
It has also been argued that in some cases, this model can be used
  to detect coding regions in linear (non-circular) DNA \cite{Yer}.

\qquad The aim of this paper is to generalize the PS model by relaxing its
main constraints, namely (i) allow the strands to be of unequal length $%
N_{1} $ and $N_{2}$ and (ii) allow the strands to be non complementary. As a
result, any base of strand $1$ can pair with any base of strand $2$
(crossings of base pairs being excluded). A preliminary account of this work
can be found in ref.\cite{cond}. In this generalized model, the
computational time scales like $N_1^{2}N_2^{2}$ but again, a Fixman-Freire
scheme brings it down to $O(N_1N_2)$. A very similar complexity reduction 
was obtained in the case of circular DNA \cite{Yer2}.
For practical purposes, this limits
the present implementation of our algorithm to sequences of up to a few kbps.

This paper is organized as follows: in section \ref{PS}, we review the
standard PS model, using partition functions \cite{Yer1} 
rather than conditional
probabilities \cite{Poland}. In section \ref{GPS}, we generalize the
recursion equations for the appropriate partition functions in order to
include the possibility of mismatches, unequal lengths and non complementary
strands. In section \ref{results}, the algorithm is first applied to
complementary strands of eukaryotic DNA of a few kbps and compared with the
standard PS algorithm: We check that there are essentially no
mismatches, thus
validating the assumptions of the PS model for complementary DNA strands of
equal length. We have then tested the algorithm with short sequences, in
order to model the hybridization of sequences of unequal lengths as it
occurs in DNA microarrays \cite{chip1,chip2}. Our simulations show that
molecular recognition is both very selective for complementary fragments of
the strands and very sensitive to single point mutations in the
sequences. Our algorithm also describes the physical process of
(complementary) sequence alignment.

\section{The Poland-Scheraga model}

\label{PS}

\subsection{Recursion relations}

Although the original Poland-Scheraga model was developped for homopolymeric
strands \cite{Pol_Scher1,Pol_Scher2,Ka_Mu_Pe}, we will focus on realistic
(heteropolymeric) DNA sequences. Exact recursion relations have been derived
by Poland \cite{Poland}, using conditional and thermodynamic probabilities.
Here, we follow an equivalent approach using partition functions, which
turns out to be easier to generalize.

We first consider two complementary strands of equal length $N$, and we
denote by $Z_f(\alpha )$ the forward partition function of
the two strands, starting at base $(1)$ and ending at base $(\alpha )$, with
bases $(\alpha )$ being paired. We model the interactions of base pairs by
stacking energies ($\varepsilon _{\alpha ,\alpha +1;\beta ,\beta +1}$),
which are known to describe nucleotides interactions in a more accurate
fashion than simple base pairing. 
%(the idea being that, in addition to the usual
%Crick-Watson pairing, there is a big component of the binding energy
%originating from the stacking of the ribose rings) 
These stacking energies account in particular 
for screened Coulomb interactions and for
hydrogen bonds between Crick-Watson pairs, and depend on pairs of adjacent
bases on the two strands, the pair $(\alpha ,\alpha +1)$ belonging to strand 
$1$ and the complementary pair $(\beta ,\beta +1)$ belonging to strand $2$
(with $\beta =\alpha $ in the PS model). Since the strands are
complementary, only $16$ stacking energies out of $4^{4}=256$ possible terms
turn out to be non zero. 
In Appendix \ref{stackingener}, we give the
values of the 10 different 
stacking energies used in the program MELTSIM
\cite{Meltsim}. These energies (which depend on the salt
concentration) will be used throughout this paper. 

\begin{figure}[htbp]
\begin{center}
\includegraphics{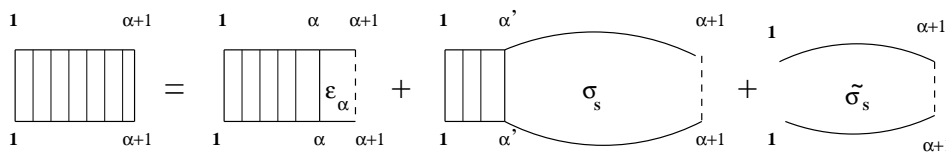}
\end{center}
\caption{Recursion relation for $Z_f(\alpha %
+1)$ (eq.(\ref{z1})) in the PS model.}
\label{f1}
\end{figure}

To find the recursion relation obeyed by $Z_f(\alpha +1)$,
we notice that there are three ways to bind a pair of chains of length $%
\alpha +1$ : Either the last pair $(\alpha ,\alpha +1)$ is stacked, or there
is a loop starting at any $(\alpha^{\prime})$ ($1\leq \alpha ^{\prime} \leq
\alpha -1$) and ending at $(\alpha +1)$, or there is no loop (Figure \ref{f1}%
).

The forward partition function therefore satisfies

\begin{equation}  \label{z1}
Z_f(\alpha +1)=e^{-\beta \varepsilon _{\alpha }}\ 
Z_f(\alpha )+\sigma _{S}\sum_{\alpha ^{\prime }=1}^{\alpha
-1}Z_f(\alpha ^{\prime }){\cal N}(2(\alpha +1-\alpha
^{\prime }))+\widetilde{\sigma}_{S}~{\cal M}(\alpha )
\end{equation}
where $\beta =1/k_{B}T$ is the inverse temperature, $\varepsilon _{\alpha
}=\varepsilon _{\alpha,\alpha +1;\alpha ,\alpha +1}$ is the stacking energy
of base pairs $(\alpha ,\alpha +1)$, $\sigma _{S}$ is the bare loop
formation (cooperativity)\ parameter and $\widetilde{\sigma }_{S}$ is the
bare free end formation parameter (we assume that these parameters are base
independent). The factor ${\cal N}(2(\alpha +1-\alpha ^{\prime }))$
counts the number of conformations of a chain starting at base $(\alpha
^{\prime })$ and ending at base $(\alpha +1)$ and is asymptotically given by 
\cite{PGG}

\begin{equation}
\label{asymp}
{\cal N}(2(\alpha +1-\alpha ^{\prime }))=\mu ^{\alpha -\alpha ^{\prime
}}f(\alpha -\alpha ^{\prime }) 
\end{equation}%
where $k_{B}\log \mu $ is the entropy per base pair and $f(x)=\frac{1}{x^{c}}
$ is the probability of return to the origin of a loop of length $2 x$. We
assume that the entropy factor $\mu $ does not depend on the chemical nature
of the base pair. The exponent $c$ depends on the interaction of the loop
with the rest of the chain: It has been extensively discussed in the context
of homopolymeric DNA \cite{Pol_Scher1,Fisher,Ka_Mu_Pe}, and is equal to $3/2 
$ for non-interacting Gaussian loops, to $\approx 1.8$ for non-interacting
self-avoiding loops, and to $\approx 2.15$ for interacting self-avoiding
loops. As stated above, eq.(\ref{asymp}) is valid only for large enough loops.
For shorter loops, one can use different formulae such as eq. (20) of ref.\cite{WB85}. 
A more accurate way to account for short loop entropies would be to
have a look-up table as is currently done in RNA folding
\cite{Tur,Vienna}. To the best of our knowledge, this has not yet been
implemented for DNA.

The last term on the r.h.s. of eq. (\ref{z1}) represents the
contribution of unbound extremities: The factor ${\cal M}(\alpha )$
counts the number of conformations of a pair of unbound chains starting at
base ($1$) and paired at base ($\alpha +1$) and is asymptotically given by 
\cite{PGG}

\begin{equation}  \label{gg}
{\cal M}(\alpha )=\mu ^{\alpha }g(\alpha )
\end{equation}
where $g(x)=\frac{1}{x^{\overline{c}}}$. For non-interacting Gaussian chains 
$\overline{c}=0$, and numerical evidence points to $\overline{c} \sim 0.09$
for self-avoiding chains \cite{Ka_Mu_Pe2}.

In a similar way, we denote by $Z_b(\alpha )$ the backward
partition function of the two strands, starting at base ($N$) and ending at
base ($\alpha $), with base ($\alpha $) being paired. To find the recursion
relation obeyed by $Z_b(\alpha )$, we again notice that there
are three ways to bind a pair of chains at base $(\alpha )$, starting from
base ($N$). The backward partition function therefore satisfies

\begin{equation}
Z_b(\alpha)=e^{-\beta \varepsilon _{\alpha }}\ 
Z_b(\alpha+1 )+\sigma _{S}\sum_{\alpha ^{\prime
}=\alpha+2}^{N}Z_b(\alpha ^{\prime }){\cal N}(2(\alpha
^{\prime}-\alpha))+\widetilde{\sigma}_{S}~{\cal M}(N-\alpha )  \label{z2}
\end{equation}

\begin{figure}[htbp]
\begin{center}
\includegraphics{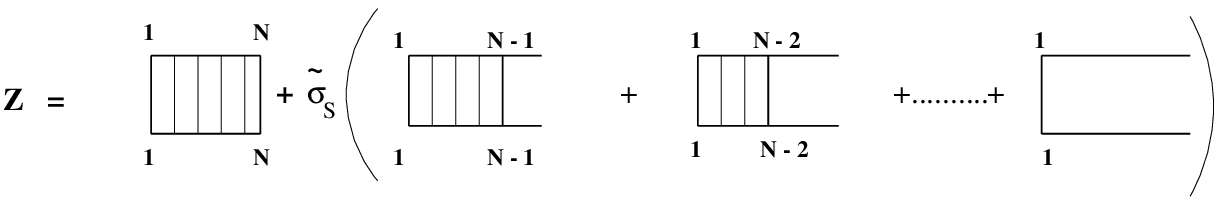}
\end{center}
\caption{Graphical representation of the thermodynamic partition function $%
\bf{\rm{Z}}$ (eq.(\protect\ref{array1})) in the PS model.}
\label{g1}
\end{figure}

The probability $p(\alpha)$ that base pair ($\alpha$) is bound can then be
expressed as

\begin{equation}
p(\alpha)=\frac{Z_f(\alpha)Z_b(\alpha)}{%
\bf{\rm{Z}}}  \label{p1}
\end{equation}%
where $\bf{\rm{Z}}$ is the thermodynamic partition function of the
two strands. Restricting ourselves to configurations with at least one bound
base pair (i.e. we do not consider dissociation), we may express $\bf{%
\rm{Z}}$ as (Figure \ref{g1})

\begin{eqnarray}  \label{array1}
\bf{\rm{Z}}&=&Z_f(N) +\widetilde{\sigma}_{S}\bigl(%
\mu g(1) Z_f(N-1)+\mu^2 g(2)Z_f(N-2)+ \cdots %
\cr &+& \mu^{N-2}g(N-2)Z_f(2)+\mu^{N-1}g(N-1)%
Z_f(1)\bigr)
\end{eqnarray}
or equivalently as

\begin{eqnarray}
\bf{\rm{Z}}&=&Z_b(1) +\widetilde{\sigma}_{S}\bigl(\mu
g(1) Z_b(2)+\mu^2 g(2)Z_b(3)+ \cdots \cr &+&
\mu^{N-2}g(N-2)Z_b(N-1)+\mu^{N-1}g(N-1)Z_b(N)%
\bigr)
\end{eqnarray}

For our purposes, we now define

\begin{equation}
Z_f^{\star }(\alpha )=\mu ^{-\alpha }Z_f%
(\alpha )
\end{equation}%
and 
\begin{equation}
Z_b^{\star }(\alpha )=\mu ^{-(N-\alpha +1)}Z_b%
(\alpha )
\end{equation}%
so that the the recursion relations read

\begin{equation}
Z_f^{\star}(\alpha +1)=e^{-\beta \varepsilon _{\alpha }-\log
\mu}\ Z_f^{\star}(\alpha )+\sigma _{0}\sum_{\alpha ^{\prime
}=1}^{\alpha -1}Z_f^{\star}(\alpha ^{\prime
})f(\alpha-\alpha ^{\prime })+\widetilde{\sigma}_{1}g(\alpha)  \label{z4}
\end{equation}%
and

\begin{equation}
Z_b^{\star }(\alpha )=e^{-\beta \varepsilon _{\alpha }-\log
\mu }\ Z_b^{\star }(\alpha +1)+\sigma _{0}\sum_{\alpha
^{\prime }=\alpha +2}^{N}Z_b^{\star }(\alpha ^{\prime
})f(\alpha ^{\prime }-\alpha -1)+\widetilde{\sigma }_{1}g(N-\alpha )
\label{z5}
\end{equation}%
where $\sigma _{0}=\frac{\sigma _{S}}{\mu }$ and $\widetilde{\sigma }_{1}=%
\frac{\widetilde{\sigma }_{S}}{\mu }$. These equations, dealing with
partition functions, are equivalent to Poland's probabilistic approach \cite%
{Poland}, as sketched in Appendix \ref{practicalps}.

Equation (\ref{p1}) now reads

\begin{equation}
p(\alpha)=\frac{Z_f^{\star}(\alpha)Z_b%
^{\star}(\alpha)}{\frac{1}{\mu} Z_f^{\star}(N)+\widetilde{%
\sigma}_1 \sum_{\alpha=1}^{N-1}Z_f^{\star}(\alpha)g(N-\alpha)%
}  \label{p2}
\end{equation}%
%
%
%or equivalently
%\begin{equation}
%p(\alpha)=\frac{Z_f^{\star}(\alpha)Z_b^{\star}(\alpha)}{\frac{1}{\mu}
%Z_b^{\star}(1)+\widetilde{\sigma}_1 \sum_{\alpha=2}^{N}Z_b^{\star}(\alpha)g(\alpha-1)} \label{p3}
%\end{equation}%

The fraction ${\theta}_{PS}$ of bound basepairs is then given by 
\begin{equation}
{\theta}_{PS}=\frac{1}{N} \sum_{\alpha=1}^N p(\alpha)
\end{equation}

This quantity can be measured by UV absorption at 268 nm \cite{WB85}. The
derivative $-d{\theta}_{PS} / dT$ with respect to temperature displays sharp
peaks at the temperatures where various fragments of the sequence open. For
a homopolymer, the fraction ${\theta}_{PS}$ is proportional to the internal
energy of the chain, and thus $-d{\theta}_{PS} / dT$ is proportional to the
specific heat. For non homogeneous sequences, it can be easily checked that
the peaks of the specific heat also coincide with those of $-d{\theta}_{PS}
/ dT$.

Since all partition functions are calculated with at least one bound pair, 
one has to include the possibility of strand dissociation 
, i.e. of two unbound strands
(Appendix \ref{dissociation}). As seen in experiments \cite{WB85}, 
strand dissociation is particularly important for
small $N$ fragments
: The corresponding calculation
of the fraction of dissociated and bound strands \cite{WB85} is given in
Appendix \ref{dissociation}. Denoting by $\theta_b$ the fraction of
bound strands, the total fraction of bound pairs (the quantity which
is measured experimentally) is given by $\theta = \theta_b
\theta_{PS}$. We will consider here only $\theta_{PS}$.
\subsection{Practical implementation}

\subsubsection{Approximation for $g(x)$}

We have mentionned above that $g(x)=x^{-\overline{c}}$. We are not aware of
any simulation with $\overline{c}\ne 0$. In order to compare our results
with previous approaches, we therefore set $\overline{c}=0$ (non-interacting
Gaussian value), so that $g(x)=1$ in eq. (\ref{z4},\ref{z5}) and (\ref{p2}).

%We have explicitly checked that
%setting $\overline{c}=0.1$ does not change the results of our
%simulations of the PS model.
The recursion relations for the PS model become

\begin{equation}
Z_f^{\star}(\alpha +1)=e^{-\beta \varepsilon _{\alpha }-\log
\mu}\ Z_f^{\star}(\alpha )+\sigma _{0}\sum_{\alpha ^{\prime
}=1}^{\alpha -1}Z_f^{\star}(\alpha ^{\prime
})f(\alpha-\alpha ^{\prime })+\widetilde{\sigma}_{1}  \label{z6}
\end{equation}%
and 
\begin{equation}
Z_b^{\star}(\alpha)=e^{-\beta \varepsilon _{\alpha }-\log
\mu}\ Z_b^{\star}(\alpha+1)+\sigma _{0}\sum_{\alpha ^{\prime
}=\alpha+2}^{N}Z_b^{\star}(\alpha ^{\prime
})f(\alpha^{\prime}-\alpha-1)+\widetilde{\sigma}_{1}  \label{z7}
\end{equation}%
where $f(x)=\frac{1}{x^c}$.

Accordingly, the probability $p(\alpha)$ that base pair ($\alpha$) is bound
is calculated as

\begin{equation}
p(\alpha)=\frac{Z_f^{\star}(\alpha)Z_b%
^{\star}(\alpha)}{\frac{1}{\mu} Z_f^{\star}(N)+\widetilde{%
\sigma}_1 \sum_{\alpha=1}^{N-1}Z_f^{\star}(\alpha)}
\label{p4}
\end{equation}
%or equivalently
%\begin{equation}
%p(\alpha)=\frac{Z_f^{\star}(\alpha)Z_b^{\star}(\alpha)}{\frac{1}{\mu}
%Z_b^{\star}(1)+\widetilde{\sigma}_1 \sum_{\alpha=2}^{N}Z_b^{\star}(\alpha)} \label{p5}
%\end{equation}
with the fraction of bound pairs ${\theta}_{PS}=\frac{1}{N}
\sum_{\alpha=1}^N p(\alpha)$.

\subsubsection{The Fixman-Freire scheme}

From a practical perspective, solving numerically equations (\ref{z6},\ref%
{z7}) requires a CPU time of order $N^2$, since one has to calculate $%
O(\alpha)$ terms for each value of $\alpha$. The Fixman-Freire (FF) method
reduces this CPU time by approximating the loop factor $f(x)$ of equations (%
\ref{z6},\ref{z7}) by 
\begin{equation}  \label{FF}
f(x)={\frac{1 }{{x}^c}} \simeq \sum_{i=1}^I a_i \ e^{-b_i x}
\end{equation}

In equation (\ref{FF}) the number ${I}$ of couples $(a_{i},b_{i})$ depends
on the desired accuracy. The parameters $(a_{i},b_{i})$ are determined by a
set of non-linear equations (see \cite{Fix_Fre}).

For a sequence of length $N=2000$, the choice ${I}=9$ gives an accuracy
better than $0.5\%$ and we have adopted this value throughout this paper.
Larger values (${I}=14$) are used in \cite{Yer} for lengths of order 150
kbps and in the program MELTSIM \cite{Meltsim}, which implements Poland's
recursion relations \cite{Poland} with a FF scheme. The CPU time of the FF
scheme scales down the computational cost from $O(N^{2})$ to $O(N\times {I})$%
, as shown by equations (\ref{z10}) and (\ref{z11}) of Appendix
\ref{practicalps}.

\subsubsection{Values of the parameters}

\label{values}

In equations (\ref{z6},\ref{z7}), one needs the values of the entropy factor 
$\mu $, the stacking energies ($\varepsilon _{\alpha }=\varepsilon _{\alpha
,\alpha +1;\alpha ,\alpha +1}$),  the exponent $c$ of the loop factor (\ref%
{FF}) and the loop formation (cooperativity) and free end formation
effective parameters $\sigma _{0}$ and $\widetilde{\sigma }_{1}$.

For complementary strands, the stacking energies we have used are the ones
of MELTSIM \cite{Meltsim}; we have also adopted the value $\log \mu=12.5047$
of this program (see Appendix \ref{stackingener}). Point mutations, when present, are
assigned a zero stacking energy : In all our numerical
calculations, we have indeed checked that the results do not depend on the
precise value of the stacking energy of the mutated pair, as
long as it is larger than half of the typical unmutated
stacking energies, i.e. $\approx -2500 ^\circ K$
.

Our calculations have been done with the Flory value $c=1.8$, and the
MELTSIM value of the cooperativity parameter $\sigma_0=1.26 \ 10^{-5}$ \cite%
{Meltsim}. As for the free end parameter $\widetilde{\sigma}_1$, we have
followed reference \cite{WB85}, and taken $\widetilde{\sigma}_1=\sqrt{%
\sigma_0} \sim 3.5 \ 10^{-3}$. 
This set of parameters will be hereafter
referred to as standard.

The exponent $c$ and cooperativity parameter $\sigma_0$ have given rise to
some discussions \cite{Ha_Me,Blo_Car}. However, we did not find in the literature
any discussion on the role 
of $\widetilde{\sigma}_1$.
This is why we have tested other values of the parameters such
as $c=2.15$, $\sigma_0=1.26 \ 10^{-4}$ and $\widetilde{\sigma}_1=1$. For the
cases studied in this paper, the changes are rather small. We find for
instance that, as long as $\widetilde{\sigma}_1$ is non-zero 
(in fact $\ge 10^{-6}$), its value is
quite irrelevant ($\widetilde{\sigma}_1=0$ corresponds to the case of paired extremities).

The boundary conditions for the recursion equations (\ref{z6}) and (\ref{z7}%
), as well as their practical implementation are exposed in Appendix
\ref{practicalps}. 

\section{Generalizing the PS model}

\label{GPS}

\subsection{Equations}

\begin{figure}[htbp]
\begin{center}
\includegraphics{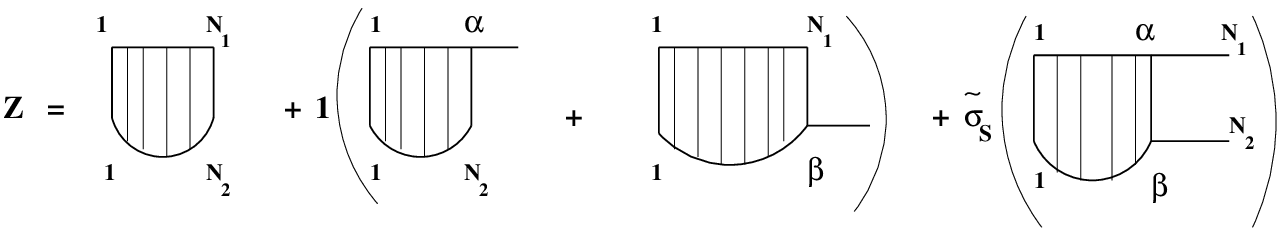}
\end{center}
\caption{Graphical representation of the thermodynamic partition function $%
\bf{\rm{Z}}$ (eq.(\protect\ref{array2})) in the GPS model.}
\label{g2}
\end{figure}

We now generalize the PS model in different ways: We allow for unequal
strand lengths denoted by $N_{1}$ and $N_{2}$, and non complementarity of
the sequences.
This in turn implies that one must allow for pairing of any base ($\alpha 
$) of strand $1$ with any other base ($\beta $) of strand $2$ (while
forbidding the crossing of base pairs) \cite{footnote}. 
Further, we allow loops (with a
factor $\sigma _{S}$), only if there is at least one unpaired base on each
strand. Finally, we associate a factor of unity, instead of $\widetilde{%
\sigma }_{S}$ for the pairing of extremities (bases ($N_{1}$) with ($\beta $%
) or ($\alpha $) with ($N_{2}$)) (see Figure \ref{g2}).

At this stage, it should be noted that since the generalized Poland-Scheraga
model (GPS) includes all configurations from the original PS model, its free
energy $F_{GPS}(T)$ is necessarily lower than that of the PS model $%
F_{PS}(T) $ 
\begin{equation}
F_{GPS}(T) \le F_{PS}(T)
\end{equation}

\begin{figure}[htbp]
\begin{center}
\includegraphics{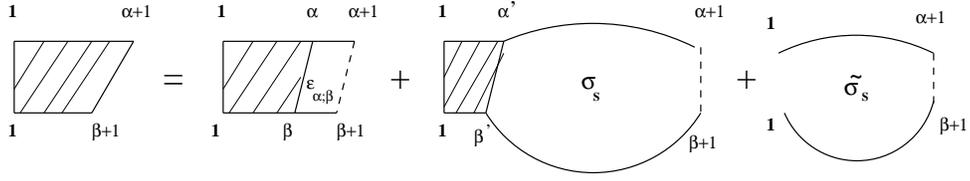}
\end{center}
\caption{Recursion relation for $Z_f(\protect\alpha%
+1,\protect\beta+1)$ (eq.(\protect\ref{partit1})) in the GPS model.}
\label{f11}
\end{figure}

We denote by $Z_f(\alpha ,\beta )$ the forward partition
function of the two strands, starting at base $(1)$ and ending respectively
at base $(\alpha )$ (strand $1$) and at base ($\beta $) (strand $2$), bases (%
$\alpha $) and $(\beta )$ being paired. We further denote by $%
Z_b(\alpha ,\beta )$ the backward partition function of the
two strands, where strand $1$ (resp. strand $2$) starts at base $(N_{1})$
(resp. $(N_{2})$) and ends at base ($\alpha )$ (resp. ($\beta $)), bases ($%
\alpha )$ and ($\beta $) being paired. Keeping the same notations as in the
PS model, and setting $\varepsilon _{\alpha ;\beta }=\varepsilon _{\alpha
,\alpha +1;\beta ,\beta +1}$, these partition functions satisfy the
recursion relations (Figure \ref{f11})

\newpage

\begin{eqnarray}  \label{partit1}
Z_f(\alpha+1,\beta+1)&=&e^{-\beta
\varepsilon_{\alpha;\beta}} Z_f(\alpha,\beta) \cr &+&
\sigma_S \sum_{\alpha^{\prime}=1}^{\alpha -1}
\sum_{\beta^{\prime}=1}^{\beta-1} Z_f(\alpha^{\prime},%
\beta^{\prime}){\cal N}(\alpha+1-\alpha^{\prime}+\beta+1-\beta^{\prime}) %
\cr &+& \widetilde{\sigma}_S{\cal M}(\alpha,\beta)
\end{eqnarray}
and 
%\newpage
\begin{eqnarray}  \label{partit2}
Z_b(\alpha,\beta)&=&e^{-\beta \varepsilon_{\alpha;\beta}} 
Z_b(\alpha+1,\beta+1) \cr &+& \sigma_S
\sum_{\alpha^{\prime}=\alpha+2}^{N_1} \sum_{\beta^{\prime}=\beta+2}^{N_2} 
Z_b(\alpha^{\prime},\beta^{\prime}){\cal N}%
(\alpha^{\prime}-\alpha+\beta^{\prime}-\beta) \cr &+& \widetilde{\sigma}_S%
{\cal M}(N_1-\alpha,N_2-\beta)
\end{eqnarray}
where ${\cal N}(x)=\mu^{\frac{x}{2}-1}f(\frac{x}{2}-1)$ and ${\cal M}%
(x,y)=\mu^{\frac{x+y}{2}}g(\frac{x+y}{2})$ . Since we have used an entropy
factor of $k_B \log \mu$ per base pair, we have assigned a factor $\frac {k_B%
}2 \log \mu$ per free base.

The probability $p(\alpha,\beta)$ that base ($\alpha$) of strand $1$ is
paired with base ($\beta$) of strand $2$ is then expressed as

\begin{equation}
p(\alpha,\beta)=\frac{Z_f(\alpha,\beta)Z_b%
(\alpha,\beta)}{\bf{\rm{Z}}}  \label{p6}
\end{equation}%
where $\bf{\rm{Z}}$ is the thermodynamic partition function of the
two strands. Equations (\ref{partit1}) and (\ref{partit2}) show that the
computational complexity of the generalized model is $O(N_1^2N_2^2)$.

As in the PS model, we take from now on $\overline{c}=0$ (i.e. $g(x)=1$).
Restricting ourselves to configurations with at least one bound base pair,
we may then express $\bf{\rm{Z}}$ as (Figure \ref{g2})

\begin{eqnarray}  \label{array2}
{\bf {\rm Z}}&=&Z_f(N_1,N_2)+\sum_{\alpha=1}^{N_1-1}\mu^{%
\frac{N_1-\alpha}{2}} Z_f(\alpha,N_2)+\sum_{\beta=1}^{N_2-1}%
\mu^{\frac{N_2-\beta}{2}}Z_f(N_1,\beta) \cr 
&+&\widetilde{%
\sigma}_{\rm S}\sum_{\alpha=1}^{N_1-1}\sum_{\beta=1}^{N_2-1}\mu^{\frac{%
N_1-\alpha+N_2-\beta}{2}}Z_f(\alpha,\beta)
\end{eqnarray}
%
%or equivalently as
%
%\begin{eqnarray}
%{\bf{\rm Z}}&=&Z_b(1,1)+\sum_{\alpha=2}^{N_1}\mu^{\frac{%
%\alpha-1}{2}}Z_b(\alpha,1)+\sum_{\beta=2}^{N_2}\mu^{\frac{%
%\beta-1}{2}}Z_b(1,\beta) \cr 
%&+&\widetilde{\sigma}_{\rm S
%}\sum_{\alpha=2}^{N_1}\sum_{\beta=2}^{N_2}\mu^{\frac{\alpha-1+\beta-1}{ 2}}%
%Z_b(\alpha,\beta)
%\end{eqnarray}
Note that the thermodynamic partition function ${\rm Z}$ can also be expressed in terms of
$Z_b(\alpha,\beta)$.

In complete analogy with the PS model, we define

\begin{eqnarray}
\label{z12}
Z_f^{\star}(\alpha,\beta)&=&\mu^{-\frac{(\alpha+\beta)}{2}} 
Z_f(\alpha,\beta) 
\end{eqnarray}
and obtain
\begin{eqnarray}  \label{z14}
Z_f^{\star }(\alpha +1,\beta +1)&=&e^{-\beta 
\varepsilon_{\alpha ;\beta }-\log \mu }\ 
Z_f^{\star}(\alpha,\beta)\cr%
&+&\sigma _{0}\sum_{\alpha ^{\prime }=1}^{\alpha -1}\sum_{\beta ^{\prime
}=1}^{\beta -1}Z_f^{\star }(\alpha ^{\prime },\beta ^{\prime
})\ f({\frac{\alpha -\alpha ^{\prime }+\beta -\beta ^{\prime }}{2}})+%
\widetilde{\sigma }_{1} 
%\cr
%Z_b^{\star}(\alpha,\beta)&=&e^{-\beta \varepsilon _{\alpha;\beta}
%-\log \mu }\ Z_b^{\star }(\alpha +1,\beta +1)\cr%
%&+&\sigma _{0}\sum_{\alpha ^{\prime }=\alpha +2}^{N_{1}}\sum_{\beta ^{\prime
%}=\beta +2}^{N_{2}}Z_b^{\star }(\alpha ^{\prime },\beta
%^{\prime })\ f({\frac{\alpha ^{\prime }-\alpha -1+\beta ^{\prime }-\beta -1}{%
%2}})+\widetilde{\sigma }_{1}
%\label{z15}
\end{eqnarray}
\medskip
We similarly define 
$Z_b^{\star}(\alpha,\beta)=\mu^{-{\frac{(N_1-\alpha+N_2-\beta)%
}{2}-1}} Z_b(\alpha,\beta)$ and get

\begin{equation}  \label{probamis}
p(\alpha,\beta)=\frac{Z_f^{\star}(\alpha,\beta)
Z_b^{\star}(\alpha,\beta)} {{\bf {\rm Z^{\star}}}}
\end{equation}
where
\begin{equation}
{\bf \rm Z^{\star}} = {\frac{1}{\mu} Z_f%
^{\star}(N_1,N_2)+\sum_{\alpha=1}^{N_1-1}Z_f%
^{\star}(\alpha,1)+\sum_{\beta=1}^{N_2-1}Z_f%
^{\star}(1,\beta) +\widetilde{\sigma}_{1}\sum_{\alpha=1}^{N_1-1}\sum_{%
\beta=1}^{N_2-1}Z_f^{\star}(\alpha,\beta)}
\end{equation}
%or equivalently
%\begin{equation}
%p(\alpha,\beta)=\frac{Z_f^{\star}(\alpha,\beta)Z_b^{\star}(\alpha,\beta)}{\frac{1}{\mu}
%Z_b^{\star}(1,1)+\sum_{\alpha=2}^{N_1}Z_b^{\star}(\alpha,1)+\sum_{\beta=2}^{N_2}Z_b^{\star}(1,\beta)
%+\widetilde{\sigma}_{1}\sum_{\alpha=2}^{N_1}\sum_{\beta=2}^{N_2}Z_b^{\star}(\alpha,\beta)}
%\end{equation}

Since we have allowed for the pairing of any $(\alpha)$ with any $(\beta)$,
we have to define the equivalent $\theta_{GPS}$ of the PS order parameter ${%
\theta}_{PS}$. For this purpose, we define for each base $(\alpha)$ of
strand 1, the base $(\beta_0(\alpha))$ of strand 2 which is maximally bound
to ($\alpha$), that is

\begin{equation}
p(\alpha,\beta_0(\alpha))=\max_{\beta =1 \cdots N_2} \bigl(p(\alpha,\beta)%
\bigr)=p_{max}(\alpha)
\end{equation}

The fraction of maximally bound pairs then reads

\begin{equation}  \label{theta2}
\theta_{GPS}=\frac{1}{N} \sum_{\alpha=1}^{N_1}p_{max}(\alpha)
\end{equation}
where $N=\min (N_1,N_2)$.

The number of mismatched pairs is defined as 
\begin{equation}  \label{mismatch}
N_M =\sum_{\alpha=1}^{N_1}\sum_{\beta=1}^{N_2} p(\alpha,\beta)-N_{GPS}
\end{equation}
where $N_{GPS} =\sum_{\alpha=1}^{N_1} p_{max}(\alpha)$.
\bigskip

As in the PS model, 
one has to include the possibility of strand dissociation 
(Appendix \ref{dissociation}). 
Denoting by $\theta _{b}$ the fraction of
bound strands, the total fraction of (maximally) bound pairs is given by $%
\theta =\theta _{b}\theta _{GPS}$. We will consider here only $\theta _{GPS}$.

\subsection{Practical implementation}

We use the same parameters and the same FF method as in the PS model. This
amounts in particular to set $f(x)=\frac{1}{x^c} \simeq \sum_{i=1}^{I}
a_i \ e^{-b_i x}$ in equations (\ref{z14}). We mostly
focus on the boundary conditions and algorithmic aspects of the resulting
recursion relations.

\subsubsection{Boundary conditions}

\label{boundcond1}

We write the list of boundary conditions relevant to the recursion relations
of equations (\ref{z14}). The corresponding strands
configurations can be easily deduced from that list. We thus have, for $%
\alpha =1,2,...N_{1}$ and $\beta =1,2,...N_{2}$ 
\begin{eqnarray*}
Z_f^{\star }(1,\beta ) &=&\mu ^{-1} \\
Z_f^{\star }(\alpha ,1) &=&\mu ^{-1} \\
Z_b^{\star }(N_{1},\beta ) &=&\mu ^{-1} \\
Z_b^{\star }(\alpha ,N_{2}) &=&\mu ^{-1}
\end{eqnarray*}

When stacking comes into play, for $\alpha =2,3,...N_{1}$ and $\beta
=2,3,...N_{2}$ we have 
\begin{eqnarray*}  \label{b2}
Z_f^{\star }(2,\beta ) &=&e^{-\beta \varepsilon _{1;\beta
-1}-\log \mu }Z_f^{\star }(1,\beta -1)+\widetilde{\sigma }%
_{1} \\
Z_f^{\star }(\alpha ,2) &=&e^{-\beta \varepsilon _{\alpha
-1;1}-\log \mu }Z_f^{\star }(\alpha -1,1)+\widetilde{\sigma }%
_{1}
\end{eqnarray*}%
and 
\begin{eqnarray*}  \label{b4}
Z_b^{\star }(N_{1}-1,\beta ) &=&e^{-\beta \varepsilon
_{N_{1}-1;\beta }-\log \mu }Z_b^{\star }(N_{1},\beta +1)+%
\widetilde{\sigma }_{1} \\
Z_b^{\star }(\alpha ,N_{2}-1) &=&e^{-\beta \varepsilon
_{\alpha ;N_{2}-1}-\log \mu }Z_b^{\star }(\alpha +1,N_{2})+%
\widetilde{\sigma }_{1}
\end{eqnarray*}%
for $\alpha =N_{1}-1,N_{1}-2,....,2,1$ and $\beta =N_{2}-1,N_{2}-2,....,2,1$.

\subsubsection{Algorithmics}
We expect the forward and backward partition functions to 
grow exponentially with the number of basepairs.
As in the PS case (see Appendix \ref{practicalps}),
we introduce recursion relations for
the logarithms (free-energy
like) of these
functions, to avoid underflows or
overflows in the computations.
Generalizing the PS approach of Appendix \ref{practicalps}.
we define
${Q}_{i}(\alpha,\beta )$ and $\mu _{i}(\alpha ,\beta )$ by 
\begin{equation}
{Q}_{i}(\alpha ,\beta )=\sum_{\alpha ^{\prime }=1}^{\alpha }\sum_{\beta
^{\prime }=1}^{\beta }Z_f^{\star }(\alpha ^{\prime },\beta
^{\prime })e^{b_{i}\frac{(\alpha ^{\prime }+\beta ^{\prime })}{2}}=e^{b_{i}%
\frac{(\alpha +\beta )}{2}}e^{\mu _{i}(\alpha ,\beta )}
\end{equation}%

We obtain 
\begin{equation}
Z_f^{\star }(\alpha ,\beta )=e^{-b_{i}\frac{(\alpha +\beta )%
}{2}}\bigl({Q}_{i}(\alpha ,\beta )+{Q}_{i}(\alpha -1,\beta -1)-{Q}%
_{i}(\alpha -1,\beta )-{Q}_{i}(\alpha ,\beta -1)\bigr)
\end{equation}%
or equivalently 
\begin{equation}
Z_f^{\star }(\alpha ,\beta )=e^{\mu _{i}(\alpha ,\beta
)}+e^{-b_{i}}e^{\mu _{i}(\alpha -1,\beta -1)}-e^{-\frac{b_{i}}{2}}\bigl(%
e^{\mu _{i}(\alpha ,\beta -1)}+e^{\mu _{i}(\alpha -1,\beta )}\bigr)
\label{mu1}
\end{equation}%
Using equation (\ref{z14}) with $\bigl(f(x)=\frac{1}{x^{c}}\simeq
\sum_{i=1}^{I}a_{i}\ e^{-b_{i}x}\bigr)$, one finally obtain a
recursion relation for the linearly growing $\mu _{i}(\alpha ,\beta
)$ 's

\begin{equation}
\mu _{i}(\alpha +1,\beta +1)=\mu _{i}(\alpha ,\beta )+\log \bigl(E+F+G+H%
\bigr)  \label{z16}
\end{equation}%
where 
\begin{eqnarray*}
E &=&-e^{-b_{i}}+e^{-\frac{b_{i}}{2}}e^{-\mu _{i}(\alpha ,\beta )}\bigl(%
e^{\mu _{i}(\alpha +1,\beta )}+e^{\mu _{i}(\alpha ,\beta +1)}\bigr) \\
F &=&e^{-\beta \varepsilon _{\alpha ;\beta }-\log \mu }\bigl(1+e^{-\mu
_{i}(\alpha ,\beta )}\bigl(e^{-b_{i}}e^{\mu _{i}(\alpha -1,\beta -1)}-e^{-%
\frac{b_{i}}{2}}\bigl(e^{\mu _{i}(\alpha ,\beta -1)}+e^{\mu _{i}(\alpha
-1,\beta )}\bigr)\bigr)\bigr) \\
G &=&\sigma _{0}\sum_{k=1}^{I}a_{k}e^{-b_{k}}e^{\mu _{k}(\alpha -1,\beta
-1)-\mu _{i}(\alpha ,\beta )} \\
H &=&\widetilde{\sigma }_{1}e^{-\mu _{i}(\alpha ,\beta )}
\end{eqnarray*}

The boundary conditions pertaining to equation (\ref{z16}) can be obtained
from section (\ref{boundcond1}), and are given in Appendix \ref{boundarygps}.

One similarly defines 
\begin{equation}
{R}_{i}(\alpha ,\beta )=\sum_{\alpha ^{\prime }=\alpha }^{N_{1}}\sum_{\beta
^{\prime }=\beta }^{N_{2}}Z_b^{\star }(\alpha ^{\prime
},\beta ^{\prime })e^{-b_{i}\frac{(\alpha ^{\prime }+\beta ^{\prime })}{2}%
}=e^{-b_{i}\frac{(\alpha +\beta )}{2}}e^{\nu _{i}(\alpha ,\beta )}
\end{equation}%
which yields 
\begin{equation}
Z_b^{\star }(\alpha ,\beta )=e^{b_{i}\frac{(\alpha +\beta )}{2%
}}\bigl({R}_{i}(\alpha ,\beta )+{R}_{i}(\alpha +1,\beta +1)-{R}_{i}(\alpha
+1,\beta )-{R}_{i}(\alpha ,\beta +1)\bigr)
\end{equation}%
or equivalently 
\begin{equation}
Z_b^{\star }(\alpha ,\beta )=e^{\nu _{i}(\alpha ,\beta
)}+e^{-b_{i}}e^{\nu _{i}(\alpha +1,\beta +1)}-e^{-\frac{b_{i}}{2}}\bigl(%
e^{\nu _{i}(\alpha ,\beta +1)}+e^{\nu _{i}(\alpha +1,\beta )}\bigr)
\label{nu1}
\end{equation}%
and using equation (\ref{z14}) with $\bigl(f(x)=\frac{1}{x^{c}}\simeq
\sum_{i=1}^{I}a_{i}\ e^{-b_{i}x}\bigr)$, one gets

\begin{equation}
\nu _{i}(\alpha ,\beta )=\nu _{i}(\alpha +1,\beta +1)+\log \bigl(E^{\prime
}+F^{\prime }+G^{\prime }+H^{\prime }\bigr)  \label{z17}
\end{equation}%
where 
\begin{eqnarray*}
E^{\prime } &=&-e^{-b_{i}}+e^{-\frac{b_{i}}{2}}e^{-\nu _{i}(\alpha +1,\beta
+1)}\bigl(e^{\nu _{i}(\alpha +1,\beta )}+e^{\nu _{i}(\alpha ,\beta +1)}\bigr)
\\
F^{\prime } &=&r\bigl(1+e^{-\nu _{i}(\alpha +1,\beta +1)}\bigl(%
e^{-b_{i}}e^{\nu _{i}(\alpha +2,\beta +2)}-e^{-\frac{b_{i}}{2}}\bigl(e^{\nu
_{i}(\alpha +2,\beta +1)}+e^{\nu _{i}(\alpha +1,\beta +2)}\bigr)\bigr)\bigr)
\\
G^{\prime } &=&\sigma _{0}\sum_{k=1}^{I}a_{k}e^{-b_{k}}e^{\nu _{k}(\alpha
+2,\beta +2)-\nu _{i}(\alpha +1,\beta +1)} \\
H^{\prime } &=&\widetilde{\sigma }_{1}e^{-\nu _{i}(\alpha +1,\beta +1)}
\end{eqnarray*}%
where $r=e^{-\beta \varepsilon _{\alpha ;\beta }-\log \mu }$

The boundary conditions pertaining to equation (\ref{z17}) can be obtained
from section (\ref{boundcond1}), and are given in Appendix \ref{boundarygps}.

Equations (\ref{z16}) and (\ref{z17}) show that the FF scheme brings down
the CPU cost from $O(N_1^2 N_2^2)$ to $O(N_1N_2 \times I)$.

The knowledge of the $\mu _{i}(\alpha ,\beta )$ and $\nu _{i}(\alpha ,\beta )
$, or of the $Z_f^{\star }(\alpha ,\beta )$ and $%
Z_b^{\star }(\alpha ,\beta )$ (through equations (\ref{mu1},%
\ref{nu1})), enables us to calculate various thermodynamical properties of
interest, including the probability $p(\alpha ,\beta )$ of pairing of bases $%
(\alpha )$ and $(\beta )$ (see eq. (\ref{probamis})), or the fraction of
(maximally) bound pairs ${\theta }_{GPS}$ (see eq. (\ref{theta2})).

\section{Results}

\label{results}

We have studied the recursion relations of the GPS model and compared them,
whenever possible, with their standard PS counterpart. In this paper, we
will present a few examples, leaving a systematic study for a future work.

\subsection{Medium length sequences}

\label{medium}

\begin{figure}[htbp]
\begin{center}
\includegraphics[height=6cm]{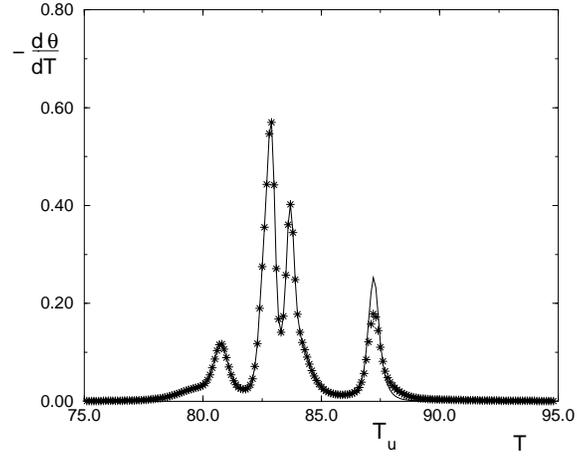}
\end{center}
\caption{The specific heat $- {\frac{d \protect\theta }{dT}}$ for
complementary sequences of length $N_1=N_2=1980$ for (i) the PS model ($\ast$%
) and (ii) the GPS model (full line). A slight difference is observed for
the final peak ($T_{u} \sim 87.2^{o}$C).}
\label{f2}
\end{figure}

In this section, we show a comparison of the two algorithms in the case of
medium length sequences ($N_1 = N_2 = N= 1980$). This sequence was extracted
from chromosome four of Drosophila melanogaster \cite{ftp}. Using the
standard set of parameters of section \ref{values}, we have first studied
the binding transition of the two complementary strands: We show in Figure %
\ref{f2} the PS and GPS results for $- {\frac{d \theta }{dT}}$ , where ${%
\theta}_{PS}=\frac{1}{N} \sum_{\alpha=1}^N p(\alpha)$ and ${\theta}_{GPS}=%
\frac{1}{N} \sum_{\alpha=1}^N p_{max}(\alpha)$.

As is clear, the two curves coincide over a wide range of temperature,
except close to the final unbinding peak (($T_{u}\sim 87.2^{o}$C), where
fluctuations allowed by the GPS model enhance the peak. 

\begin{figure}[htbp]
\begin{center}
\includegraphics[height=6cm]{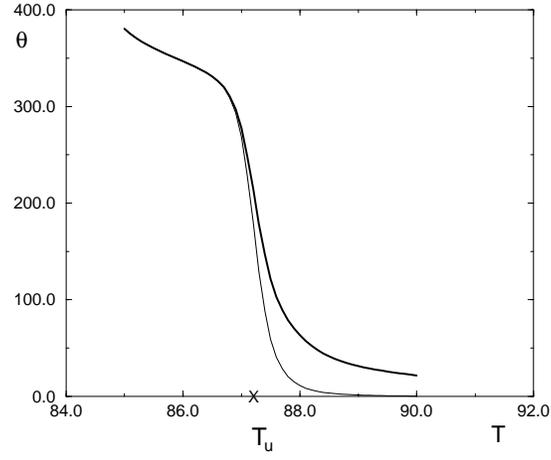}
\end{center}
\caption{Number of bound pairs for the PS (thick line) and for the GPS
  (thin line) around $T_u$.
}
\label{f22}
\end{figure}

To illustrate this point, we plot the total number of bound pairs 
$N \theta_{PS}$ and
$N \theta_{GPS}$ around the
unbinding temperature $T_u \sim 87.2^\circ$C
(Figure \ref{f22}): For instance at $88 ^\circ$C, the
PS model overestimates the total number of bound pairs by about 60.

In addition, a study of the
number $N_{M}$ of mismatches as a function of temperature indeed shows that $%
N_{M}\sim 0,\ T<T_{u}$. Since the partition function $\bf{\rm{Z}}$
includes configurations with at least one bound base pair, one finds $%
N_{M}\sim 1,\ T>T_{u}$.

We have studied several sequences with the same results, thereby validating
the hypothesis that DNA denaturation can be modeled by the PS model ($\alpha
=\beta $), implying a very strong selectivity of molecular recognition.

\begin{figure}[htbp]
\begin{center}
\includegraphics[height=6cm]{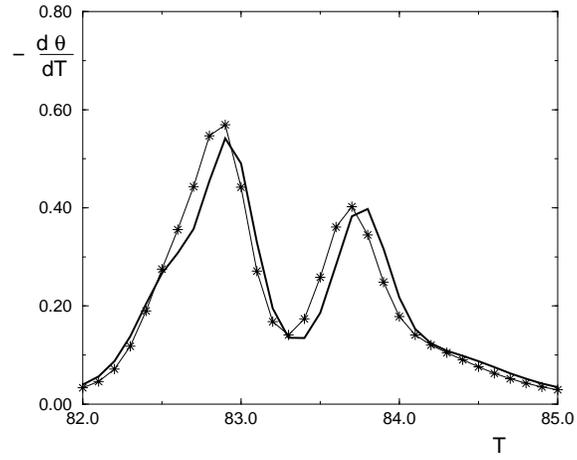}
\end{center}
\caption{Blow-up of the two central peaks of Figure \protect\ref{f2} for the
GPS model (i) standard parameters ($\ast$) (ii) parameters of reference 
\protect\cite{Blo_Car} (thick line).}
\label{f3}
\end{figure}

\begin{figure}[htbp]
\begin{center}
\includegraphics[height=6cm]{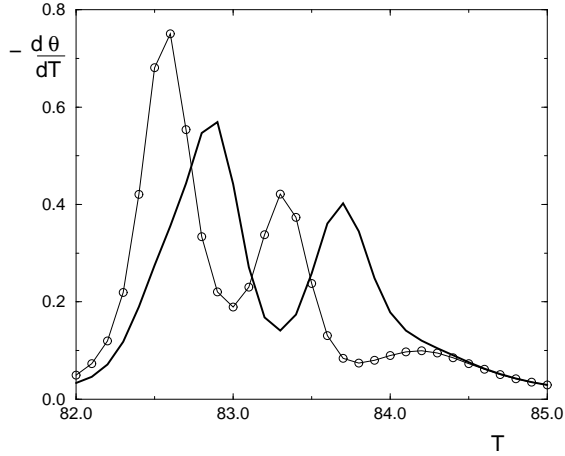}
\end{center}
\caption{Blow-up of the two central peaks for the GPS model (i)
complementary strands (thick line) (ii) a mutation in the middle of strand 1
($\circ$).}
\label{f4}
\end{figure}

We have further considered the influence of the ($c,\sigma_0,\widetilde{%
\sigma}_1=\sqrt{\sigma_0}$) parameters in the framework of the GPS model.
Following a recent proposal \cite{Blo_Car}, we have compared the standard
set of parameters with the set ($c=2.15,\sigma_0=1.26 \ 10^{-4},\widetilde{%
\sigma}_1=\sqrt{\sigma_0}$).

The results are shown in Figure \ref{f3}. As in the standard PS model, the
dependence of the results on these parameters is very weak, in agreement
with reference \cite{Blo_Car}.

Finally, we have considered the influence of a point mutation on strand 1
within the GPS model, and using standard
parameters. The corresponding
stacking energies were set to zero. In Figure \ref{f4}, we show the effect
of single point mutation in the middle region of strand 1: it slightly
shifts the whole curve towards lower temperatures. The effect of multiple
mutations and the effect of their location on the strands will be studied in
a future work.

\subsection{Hybridization of short fragments and the influence of mutations}
As mentioned in the introduction, the hybridization of short DNA fragments
is of interest for DNA microarrays. We have compared the PS and GPS
models for complementary fragments of identical lengths
$N_1=N_2=30$. In that case, there is only one peak in $-\frac
{d\theta} {dT}$; around this peak, the situation is very similar to
that observed around the last peak $T_u$ of the previous section. The
PS model overestimates the number of bound pairs by about 5 basepairs.

\begin{figure}[htbp]
\begin{center}
\includegraphics{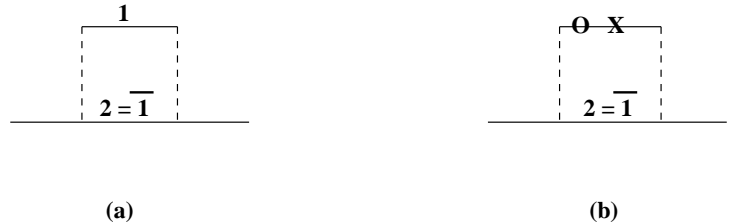}
\end{center}
\caption{Recognition of two strands of different lengths ($N_1=30, \ N_2=70$%
) (a) Fragment $\overline{1}$ of strand 2 is complementary to strand 1 (b)
One creates a mutation on strand 1, either in the middle ($\bf{X}$, $%
\protect\alpha=15$) or at the end ($\bf{O}$, $\protect\alpha=5$).}
\label{f5}
\end{figure}

Using the GPS model, we have also
studied the hybridization of strand 1 ($N_1=30$) taken from the same
drosophile chromosome \cite{ftp}, with a fragment of length $N_2=70$
containing the complementary of strand 1 in its middle section (Figure \ref%
{f5}(a). The first and the last 20 bases of strand 2 are taken from a
different region of the same DNA fragment. We have then studied the same
system, with a point mutation (see Figure \ref{f5}(b)), in the middle of
strand 1 ($\bf{X}$) or close to its extremities ($\bf{O}$).

\begin{figure}[htbp]
\begin{center}
\includegraphics[height=6cm]{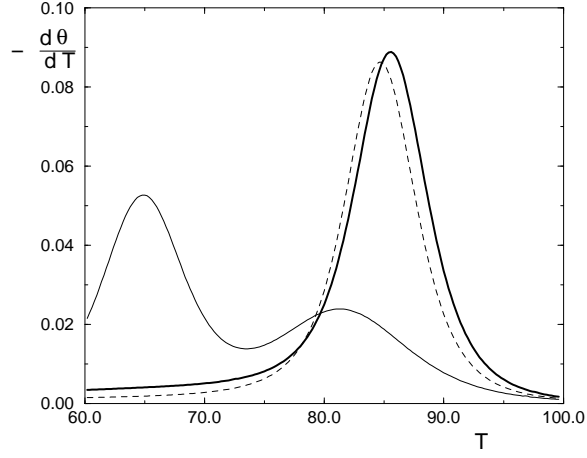}
\end{center}
\caption{The effect of mutations $\bf{X}$ (thin line) and $\bf{O}$
(dashed line) on the no mutation situation (thick line).}
\label{f6}
\end{figure}

\begin{figure}[htbp]
\begin{center}
\includegraphics[height=6cm]{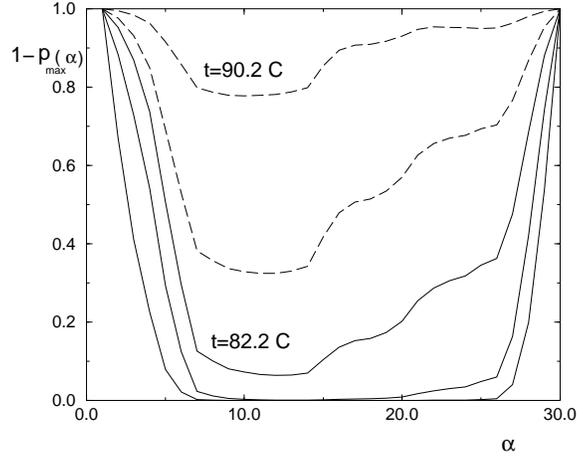}
\end{center}
\caption{Opening probability $(1-p_{max}(\protect\alpha))$ along
strand 1 for temperatures t=60, 74.2, 82.2, 85.6, 90.2, for the no mutation
case.}
\label{f7}
\end{figure}

\begin{figure}[tbp]
\begin{center}
\includegraphics[height=6cm]{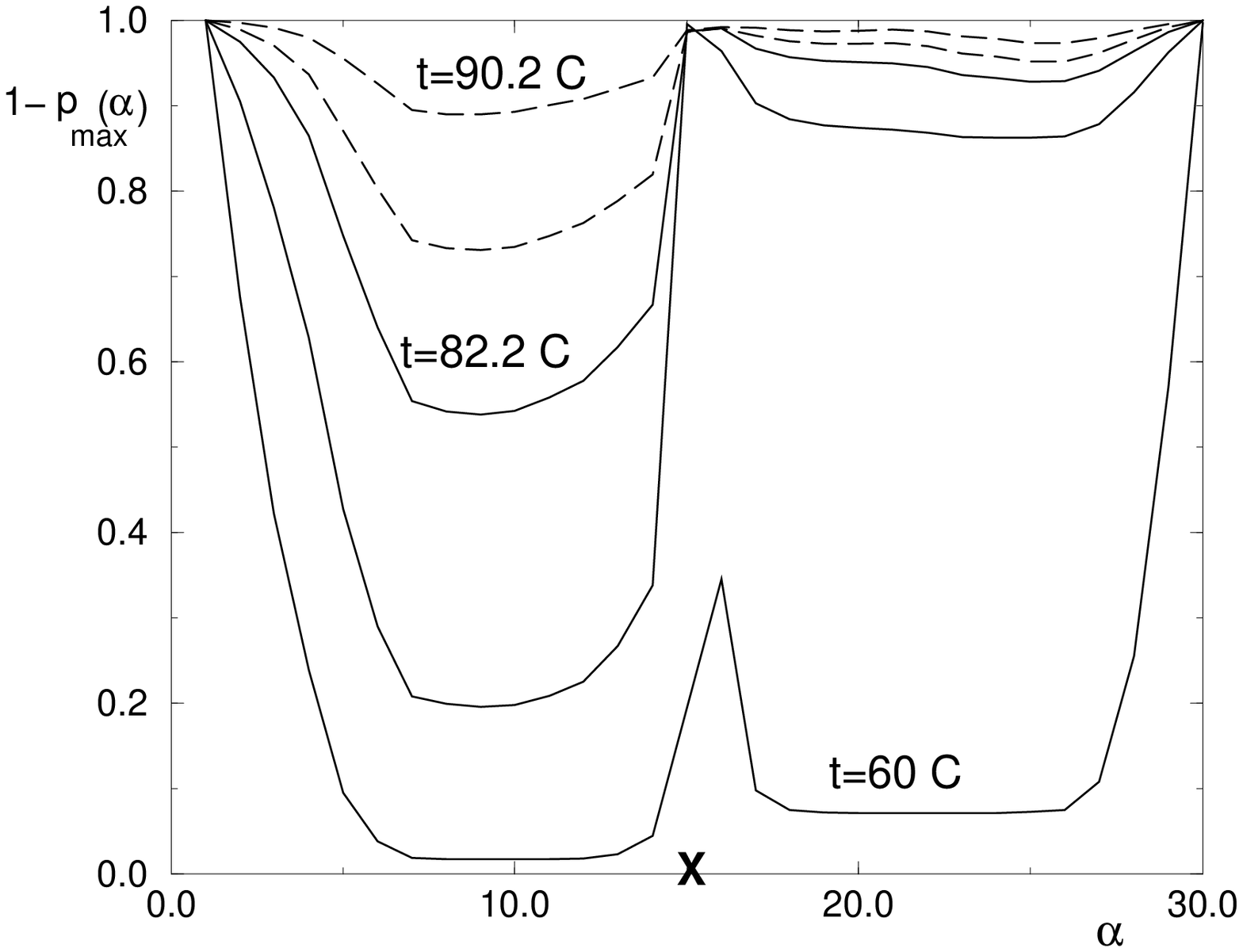}
\end{center}
\caption{Opening probability $(1-p_{max}(\protect\alpha))$ along
strand 1 for temperatures t=60, 74.2, 82.2, 85.6, 90.2, for mutation ($%
\bf{X}$) at $\protect\alpha=15$.}
\label{f8}
\end{figure}

\begin{figure}[htp]
\begin{center}
\includegraphics[height=6cm]{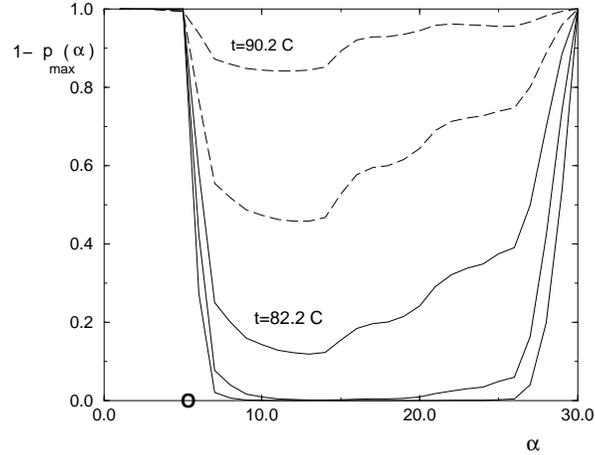}
\end{center}
\caption{Opening probability $(1-p_{max}(\protect\alpha))$ along
strand 1 for temperatures t=60, 74.2, 82.2, 85.6, 90.2, for mutation ($%
\bf{O}$) at $\protect\alpha=5$.}
\label{f9}
\end{figure}

In Figure \ref{f6}, we plot $- {\frac{d \theta }{dT}}$ for the three cases
mentioned above. For short sequences, the effect of mutation ($\bf{X}$)
in the middle section of the strand is important: the curve has two maxima
instead of one, corresponding to the opening of the two subfragments of the
strand. The effect of mutation ($\bf{O}$) is much weaker since it is
located near the extremity of strand 1. This general feature has been
checked on many different choices of strand 1. The physical origin of this
phenomenon is easy to understand: since $\widetilde{\sigma}_1 \ne 0$,
fluctuations are larger near the ends of the strands, and since the
extremities of strand 1 are nearly molten, the effect of a mutation in this
region is very weak.
In Figures (\ref{f7},\ref{f8},\ref{f9}), we plot the opening
probability $\bigl(1-p_{max}(\alpha)\bigr)$ along strand 1, for various
temperatures and for the cases of no mutation, mutation ($\bf{X}$), and
mutation ($\bf{O}$).

As stated in section \ref{PS}, the
study of short fragments and loops, would ideally require the use
of look-up tables for the entropies, rather than the asymptotic
formula (\ref{asymp}). In addition, we have checked that our results
do not depend on the
precise value of the stacking energy of the mutated pair, as
long as it is larger than half of the typical unmutated
stacking energies, i.e. $\approx -2500 ^\circ K$.

\newpage
\section{Conclusion}
\begin{figure}[htbp]
\begin{center}
\includegraphics[height=6cm]{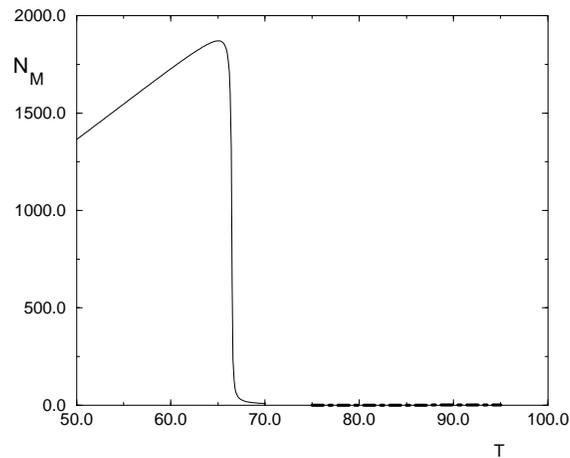}
\end{center}
\caption{Number of mismatched pairs $N_M$ (eq. (\protect\ref{mismatch})) as
a function of temperature for homopolymeric (aaaa...) and (tttt...) strands
of length $N=1980$. For comparison, the heteropolymeric case studied in
section \protect\ref{medium} is shown as the heavy dot-dashed line, lying
close to the temperature axis. }
\label{f10}
\end{figure}

We have presented a generalization of the Poland-Scheraga model, which
allows for the pairing of any bases on two DNA strands of unequal lengths.
The resulting algorithm has been implemented within a Fixman-Freire scheme.
The examples that we have given emphasize the fact that, for realistic
sequences, the Poland-Scheraga model captures the essence of the DNA strands
binding selectivity. In this respect, the generalized model is particularly useful
in certain specific situations (such as 
tandem repeats \cite{repeat}). To further illustrate the strong selectivity of real sequences, 
we compare the number $N_M$ of mismatches, as given by equation (\ref%
{mismatch}), for homopolymeric and realistic complementary strands of length $N=1980$
(using the standard set of parameters) in Figure \ref{f10}.

Our approach is also useful for strands of unequal lengths, in particular
for DNA microarrays. The role of mutations remains to be systematically
studied. 

\begin{acknowledgments}
We wish to thank E. Yeramian for many useful discussions.
\end{acknowledgments}

\newpage

\appendix
\section{Dissociation equilibrium of double stranded DNA}
\label{dissociation}

In this appendix, we show how the thermodynamic fraction of dissociated DNA
chains can be computed. Consider a solution of $M$ single strands of DNA
(denoted S$_{1}$) of size $N_{1}$ and $M$ single strands of DNA (denoted S$%
_{2}$) of size $N_{2}$ in a volume $V$. The total concentration of single
strands in the solution is

\[
c_{T}=\frac{2M}{V} 
\]

These strands can associate (or hybridize) into double stranded DNA (denoted
D) according to the chemical reaction

\centerline{D $\rightleftharpoons$ S$_1$ + S$_2$}

We denote by $z_{1}$ the partition function of a single strand S$_{1}$ and
by $z_{2}$ that of a single strand S$_{2}$ both with fixed origin. In a
purely entropic model for the single strands, these partition function would
read

\[
z_{1,2}=\mu ^{N_{1,2}/2}N_{1,2}^{\gamma -1} 
\]%
but of course, one can use more realistic models.

Let $z$ denote the partition function of two hybridized strands D (with
fixed origin), i.e. strands with at least one pair of bound bases. This last
partition function $z$ is nothing but the total partition function $\bf{%
\rm{Z}}$ calculated in section \ref{PS}.

A generic configuration of the solution consists of $p$ hybridized strands
D, $M-p$ single strands S$_{1}$ and $M-p$ single strands S$_{2}$. The
partition function of the system can thus be written as

\begin{equation}
Z=\sum_{p=0}^{M}\frac{\left( z_{1}z_{2}\right) ^{M-p}~z^{p}}{\left(
M-p\right) !^{2}~p!}~V^{2M-p}  \label{zsol}
\end{equation}

In the thermodynamic limit ($M\rightarrow \infty $) the sum in eq. (\ref%
{zsol}) is dominated by the value of $p$ which maximizes the generic term.
Using the Stirling formula, the equation for $p$ reads

\begin{equation}
\frac{\left( M-p\right) ^{2}}{pV}=\frac{z_{1}z_{2}}{z}  \label{saddle}
\end{equation}%
Introducing the dissociation constant $K$ of the Law of Mass Action

\begin{eqnarray}
K &=&\frac{[\rm{S}_{1}].[\rm{S}_{2}]}{[\rm{D}]}  \nonumber \\
&=&\frac{\left( M-p\right) ^{2}}{pV}  \label{mass1}
\end{eqnarray}%
we have

\[
K=\frac{z_{1}z_{2}}{z} 
\]

Denoting by $\theta _{b}$ the fraction of bound strands

\[
\theta _{b}=\frac{p}{M} 
\]%
equation (\ref{saddle}) is equivalent to the quadratic equation

\begin{equation}
\theta _{b}^{2}-2\left( 1+\frac{K}{c_{T}}\right) \theta _{b}+1=0  \label{eq1}
\end{equation}%
which implies

\begin{equation}
\theta _{b}=1+\frac{K}{c_{T}}-\sqrt{\left( \frac{K}{c_{T}}\right) ^{2}+2%
\frac{K}{c_{T}}}
\end{equation}

Note that there is a factor 4 in our eq. (\ref{eq1}) compared to eq. (34) of
ref.\cite{WB85} due to the fact that we consider the two strands as distinct
objects. %\newpage

\section{Practical implementation of the PS model}
\label{practicalps}

\subsection{Boundary conditions for the PS model}

Equation (\ref{z6}) applies for $\alpha \ge 2$, and equation (\ref{z7}) for $%
\alpha \le N-2$. We must initiate the recursion relations with boundary
conditions on $Z_f^{\star}(m)$ for $m=1,2$, and on $%
Z_b^{\star}(m)$ for $m=N,N-1$. A similar situation exists in
Poland's probabilistic approach \cite{Poland}.

Since $\widetilde{\sigma}_1 \ne 0$, these boundary conditions are important.
We have 
\begin{eqnarray}
Z_f(1)&=&1 \cr Z_f(2)&=&e^{-\beta
\varepsilon_{1}} Z_f(1) + \widetilde{\sigma}_{S}\mu
\end{eqnarray}
which gives

\begin{eqnarray}
Z_f^{\star}(1)&=&\mu^{-1} \cr Z_f%
^{\star}(2)&=&e^{-\beta \varepsilon _{1}-\log \mu} Z_f%
^{\star}(1) +\widetilde{\sigma}_1
\end{eqnarray}

Similarly, we have 
\begin{eqnarray}
Z_b(N)&=&1 \cr Z_b(N-1)&=&e^{-\beta
\varepsilon _{N-1}} Z_b(N) +\widetilde{\sigma}_{S}\mu
\end{eqnarray}

yielding 
\begin{eqnarray}
Z_b^{\star}(N)&=&\mu^{-1} \cr Z_b%
^{\star}(N-1)&=&e^{-\beta \varepsilon _{N-1}-\log \mu} Z_b%
^{\star}(N) +\widetilde{\sigma}_1
\end{eqnarray}

\subsection{Algorithmics}

Inserting equation (\ref{FF}) in (\ref{z6}), we have

\begin{equation}
Z_f^{\star}(\alpha +1)=e^{-\beta \varepsilon _{\alpha }-\log
\mu}\ Z_f^{\star}(\alpha )+\sigma _{0}\sum_{i=1}^{I}
a_i\sum_{\alpha ^{\prime }=1}^{\alpha -1} \ e^{-b_i(\alpha-\alpha ^{\prime
}) }Z_f^{\star}(\alpha ^{\prime })+\widetilde{\sigma}_{1}
\label{z8}
\end{equation}

As expected, and made explicit in equation (\ref{z8}), the growth of
$Z_f^{\star}(\alpha)$ is exponential with $\alpha$. To 
deal with (free-energy like) quantities that are linear in $\alpha$,
we define ${Q}_i(\alpha)$ and $\mu_i(\alpha)$ by 
\begin{equation}
{Q}_i(\alpha)=\sum_{\alpha ^{\prime }=1}^{\alpha}Z_f%
^{\star}(\alpha ^{\prime })e^{b_i\alpha^{\prime}}=e^{b_i\alpha}
e^{\mu_i(\alpha)}
\end{equation}

and get

\begin{equation}
Z_f^{\star}(\alpha)=e^{-b_i\alpha}\bigl({Q}_i(\alpha)-{Q}%
_i(\alpha-1)\bigr)=e^{\mu_i(\alpha)}-e^{-b_i}e^{\mu_i(\alpha-1)}
\end{equation}

Using equation (\ref{z8}), one finally gets a recursion relation on
the (linearly growing) $\mu_i(\alpha)$ 's
\begin{equation}
\mu_i(\alpha+1)=\mu_i(\alpha)+\log \bigl(A+B+C+D\bigr)  \label{z10}
\end{equation}
where 
\begin{eqnarray}
A&=&e^{-b_i} \cr B&=&e^{-\beta \varepsilon_{\alpha }-\log
\mu}(1-e^{-b_i}e^{\mu_i(\alpha-1)-\mu_i(\alpha)}) \cr C&=&\sigma_0
\sum_{k=1}^{I}a_k e^{-b_k}e^{\mu_k(\alpha-1)-\mu_i(\alpha)} \cr D&=&%
\widetilde{\sigma}_1 e^{-\mu_i(\alpha)}
\end{eqnarray}

Equation (\ref{z10}) is to be iterated with boundary conditions 
\begin{eqnarray}
\mu_i(1)&=&\log Z_f^{\star}(1)=-\log \mu \cr \mu_i(2)&=&\log
(e^{-b_i}Z_f^{\star}(1)+Z_f^{\star}(2))
\end{eqnarray}

One similarly defines ${R}_i(\alpha)$ and $\nu_i(\alpha)$ through 
\begin{equation}
{R}_i(\alpha)=\sum_{\alpha ^{\prime }=1}^{\alpha}Z_b%
^{\star}(\alpha ^{\prime })e^{-b_i\alpha^{\prime}}=e^{-b_i\alpha}
e^{\nu_i(\alpha)}
\end{equation}
which imply 
\begin{equation}
Z_b^{\star}(\alpha)=e^{b_i\alpha}\bigl({R}_i(\alpha)-{R}%
_i(\alpha+1)\bigr)=e^{\nu_i(\alpha)}-e^{-b_i}e^{\nu_i(\alpha+1)}
\end{equation}
and

\begin{equation}
\nu_i(\alpha)=\nu_i(\alpha+1)+\log \bigl(A^{\prime}+B^{\prime}+C^{%
\prime}+D^{\prime}\bigr)  \label{z11}
\end{equation}
where 
\begin{eqnarray}
A^{\prime}&=&e^{-b_i} \cr B^{\prime}&=&e^{-\beta \varepsilon_{\alpha }-\log
\mu}(1-e^{-b_i}e^{\nu_i(\alpha+2)-\nu_i(\alpha+1)}) \cr C^{\prime}&=&%
\sigma_0 \sum_{k=1}^{I}a_k e^{-b_k}e^{\nu_k(\alpha+2)-\nu_i(\alpha+1)} \cr %
D^{\prime}&=&\widetilde{\sigma}_1 e^{-\nu_i(\alpha+1)}
\end{eqnarray}
with corresponding boundary conditions 
\begin{eqnarray}
\nu_i(N) &=& \log Z_b^{\star}(N)=-\log \mu \cr \nu_i(N-1) &=&
\log (e^{-b_i}Z_b^{\star}(N)+Z_b^{\star}(N-1))
\end{eqnarray}

The knowledge of the $\mu_i(\alpha)$'s and of the $\nu_i(\alpha)$'s, that is
of the $Z_b^{\star}(\alpha)$'s and of the $Z_b%
^{\star}(\alpha)$'s, enables us to calculate various thermodynamical
properties of interest, including the probability $p(\alpha)$ of pairing of
basepair $(\alpha)$ (see eq. (\ref{p4})), or the fraction of bound pairs ${%
\theta}_{PS}$.

\subsection{ Connection with Poland's approach}

Poland \cite{Poland} has derived recursion relations using conditional and
thermodynamic probabilities. The link with the present approach is made
clear, if one rewrites eq (\ref{z2}) as

\begin{equation}
e^{-\beta \varepsilon _{\alpha }}\ \frac{Z_b(\alpha+1)}{%
Z_b(\alpha)}+\sigma _{S}\sum_{\alpha ^{\prime }=\alpha+2}^{N}%
\frac{Z_b(\alpha^{\prime})}{Z_b(\alpha)}%
{\cal N}(2(\alpha ^{\prime}-\alpha))+\widetilde{\sigma}_{S}~\frac{%
{\cal M}(N-\alpha)}{Z_b(\alpha)}=1  \label{z3}
\end{equation}

Equation (\ref{z3}) is term by term identical to Poland's equation (1b) on
conditional probabilities. Similarly, inserting equation (\ref{z1}) in
equation (\ref{p1}) leads to Poland's equation (10) on thermodynamic
probabilities \cite{Poland}.

\section{ Boundary conditions for the generalized model}
\label{boundarygps}

The boundary conditions for the $\mu_i(\alpha,\beta)$'s corresponding to
equation (\ref{z16}) can easily be found from section (\ref{boundcond1}) and
from equation (\ref{mu1}) as

\begin{equation}
\mu_i(\alpha,1)=-\log \mu+\log \frac{1-e^{-\frac{b_i}{2}\alpha}}{1-e^{-\frac{%
b_i}{2}}}
\end{equation}
for $\alpha=1,2,\cdots, N_1$. and 
\begin{equation}
\mu_i(1,\beta)=-\log \mu+\log \frac{1-e^{-\frac{b_i}{2}\beta}}{1-e^{-\frac{%
b_i}{2}}}
\end{equation}
for $\beta=1,2,\cdots, N_2$.

We also have 
\begin{equation}
\mu_i(\alpha,2)=-\log \mu+\log \bigl( \ \frac{1-e^{-\frac{b_i}{2}\alpha}}{e^{%
\frac{b_i}{2}}-1}+ \sum_{\alpha^{\prime}=1}^{\alpha}e^{\frac {b_i}{2}
(\alpha^{\prime}-\alpha)} Z_f^{\star}(\alpha^{\prime},2) \ %
\bigr)
\end{equation}
for $\alpha=2,3,\cdots,N_1$, and 
\begin{equation}
\mu_i(2,\beta)=-\log \mu+\log \bigl( \ \frac{1-e^{-\frac{b_i}{2}\beta}}{e^{%
\frac{b_i}{2}}-1}+ \sum_{\beta^{\prime}=1}^{\beta}e^{\frac {b_i}{2}
(\beta^{\prime}-\beta)} Z_f^{\star}(2,\beta^{\prime}) \ %
\bigr)
\end{equation}
for $\beta=2,3,\cdots,N_2$.

The boundary conditions for the $\nu_i(\alpha,\beta)$'s corresponding to
equation (\ref{z17}) can easily be found from section (\ref{boundcond1}) and
from equation (\ref{nu1}) as 
\begin{equation}
\nu_i(\alpha,N_2)=-\log \mu+\log \frac{1-e^{-\frac{b_i}{2}(N_1-\alpha+1)}}{%
1-e^{-\frac{b_i}{2}}}
\end{equation}
for $\alpha=1,2,\cdots, N_1$. and 
\begin{equation}
\nu_i(N_1,\beta)=-\log \mu+\log \frac{1-e^{-\frac{b_i}{2}(N_2-\beta+1)}}{%
1-e^{-\frac{b_i}{2}}}
\end{equation}
for $\beta=1,2,\cdots, N_2$.

We also have 
\begin{equation}
\nu_i(\alpha,N_2-1)=-\log \mu+\log \bigl( \ \frac{1-e^{-\frac{b_i}{2}%
(N_1-\alpha+1)}}{e^{\frac{b_i}{2}}-1}+
\sum_{\alpha^{\prime}=\alpha}^{N_1}e^{-\frac {b_i}{2} (\alpha^{\prime}-%
\alpha)} Z_b^{\star}(\alpha^{\prime},N_2-1) \ \bigr)
\end{equation}
for $\alpha=1,2,\cdots,N_1-1$. and 
\begin{equation}
\nu_i(N_1-1,\beta)=-\log \mu+\log \bigl( \ \frac{1-e^{-\frac{b_i}{2}%
(N_2-\beta+1)}}{e^{\frac{b_i}{2}}-1}+ \sum_{\beta^{\prime}=\beta}^{N_2}e^{-%
\frac {b_i}{2} (\beta^{\prime}-\beta)} Z_b^{\star}(N_1-1,%
\beta^{\prime}) \ \bigr)
\end{equation}
for $\beta=1,2,,\cdots,N_2-1$.

\section{The MELTSIM stacking energies}
\label{stackingener}
For complementary base pairs, the stacking energies we use are the ones of
the program MELTSIM. They are written as $\varepsilon_{\alpha ,\alpha
+1;\beta ,\beta +1}=-12.5047 \ T \bigl(n(\alpha),n(\alpha+1)\bigr)$, where
$n(\alpha)$ denotes the chemical nature (A,T,G,C) of base
($\alpha$), and where $T\bigl(n(\alpha),n(\alpha+1)\bigr)$ has the
dimension of a temperature. The factor $12.5047$ is the entropy loss
due to the formation of a base pair divided by Boltzmann's constant
($12.5047=\frac{24.85}{1.99}$). Bases ($\alpha$) and ($\alpha+1$)
belong to strand 1, identified as going from 5' to 3'. If $s$ denotes
the salt concentration, the effective temperatures
$T\bigl(n(\alpha),n(\alpha+1)\bigr)$ are generically given as  
\begin{equation*}
T\bigl(n(\alpha),n(\alpha+1)\bigr)=a_0\bigl(n(\alpha),n(\alpha+1)\bigr)
\log s+ b_0\bigl(n(\alpha),n(\alpha+1)\bigr)
\end{equation*}
where $a_0$ and $b_0$ do not depend on $s$. In our simulations, we took $%
s=0.0745$. For this particular value, we list below the effective
temperatures $T(n(\alpha),n(\alpha+1))$ with A $=\bf{1}$, T $=\bf{3}$,
G $=\bf{2}$, C $=\bf{4}$. 

\begin{eqnarray*}
T(\bf{1},\bf{1})&=&T({\bf{3,3}})= 339.68 \ K \cr
T(\bf{1},\bf{2})&=&T({\bf{4,3}})= 353.32 \ K \cr
T(\bf{1},\bf{3})&=& 341.72 \ K \cr
T(\bf{1},\bf{4})&=&T({\bf{2,3}})= 378.83 \ K \cr
T(\bf{2},\bf{1})&=&T({\bf{3,4}})= 357.96 \ K \cr
T(\bf{2},\bf{2})&=&T({\bf{4,4}})= 372.73 \ K \cr
T(\bf{2},\bf{4})&=& 408.99 \ K \cr
T(\bf{3},\bf{1})&=& 326.65 \ K \cr
T(\bf{3},\bf{2})&=&T({\bf{4,1}})= 341.30 \ K \cr
T(\bf{4},\bf{2})&=& 361.49 \ K \cr
\end{eqnarray*}

For non complementary bases (e.g. in the case of point mutations), we take %
$\varepsilon_{\alpha,\alpha +1;\beta ,\beta +1}=0$.


\newpage

\begin{thebibliography}{99}
\bibitem{CW}
J.D. Watson and F.H.C. Crick, Nature, {\bf 171}, 737 (1953).

\bibitem{Pol_Scher2} D. Poland and H.A. Scheraga eds., \textit{Theory of
Helix-Coil transition in Biopolymers}, Academic Press, New York (1970).

\bibitem{chip1} S.P.A. Fodor, Science, \textbf{277}, 393 (1997).
\bibitem{chip2} E. M. Southern, Methods Mol. Biol., \textbf{170}, 181 (2001).

\bibitem{Pol_Scher1} D. Poland and H.A. Scheraga, J. Chem. Phys., \textbf{45}%
, 1456, 1464 (1966).

\bibitem{Poland} D. Poland, Biopolymers, \textbf{13}, 1859 (1974).

\bibitem{Poland2} D. Poland, Biopolymers, \textbf{73}, 216 (2004).

\bibitem{Fix_Fre} M. Fixman and J.J. Freire, Biopolymers, \textbf{16}, 2693
(1977)

\bibitem{Yer1} E. Yeramian et al., Biopolymers, \textbf{30}, 481 (1990).

\bibitem{Meltsim} R.D. Blake, J.W. Bizarro, J.D. Blake, G.R. Day, S.G.
Delcourt, J. Knowles, K.A. Marx and J. SantaLucia Jr., Bioinformatics, 
\textbf{15}, 370-375 (1999); see also J. Santalucia Jr., Proc. Natl. Acad.
Sci. USA, \textbf{95}, 1460 (1998).

\bibitem{WB85} R.M. Wartell and A.S. Benight, Phys. Repts., \textbf{126}, 67
(1985).

\bibitem{Yer} E. Yeramian, Genes, \textbf{255}, 139, 151 (2000).

\bibitem{cond} T. Garel and H. Orland, cond-mat/ 0304080.

\bibitem{Yer2} E. Yeramian, Europhys. Lett., \textbf{25}, 49 (1994).

\bibitem{Ka_Mu_Pe} Y. Kafri, D. Mukamel and L. Peliti, Phys. Rev. Lett., 
\textbf{85}, 4988 (2000).

\bibitem{PGG} P.G. de Gennes, \textit{Scaling concepts in polymer physics} ,
Cornell University Press, Ithaca, (1979).

\bibitem{Fisher} M.E. Fisher, J. Chem. Phys., \textbf{45}, 1469 (1966).

\bibitem{Tur} 
D.H. Mathews, J. Sabina, M. Zucker and H. Turner,
J. Mol. Biol., \textbf{288}, 911 (1999). 

\bibitem{Vienna}Ivo L. Hofacker, Walter Fontana, Peter F. Stadler,
L. Sebastian Bonhoeffer, Manfred Tacker, and Peter Schuster, 
Monatsh.Chem., \textbf{125}, 167 (1994).

\bibitem{Ka_Mu_Pe2} Y. Kafri, D. Mukamel and L. Peliti, Eur. Phys. J. B, 
\textbf{27}, 135 (2002).

\bibitem{Ha_Me} A. Hanke and R. Metzler, Phys. Rev. Lett., \textbf{90},
159801 (2003); Y. Kafri, D. Mukamel and L. Peliti, ibidem, 159802 (2003).

\bibitem{Blo_Car} R. Blossey and E. Carlon, Phys. Rev. E, \textbf{68},
061911 (2003) and references therein.

\bibitem{footnote}
In the homopolymeric case with $N_{1}=N_{2}$ and stacking energies
independent of the sequence, allowing for mismatches leads to an effective
PS model where the loop exponent $c$ is replaced by $c-1$ \cite{Pol_Scher2}.

\bibitem{ftp} For the medium length sequences, we have taken as strand 1,
the first 33 lines of AE014135 (Drosophila melanogaster chromosome 4) at
http://www.ebi.ac.uk/genomes/eukaryota.html. For the short sequences
analysis, we have extracted ten to twenty sequences of the same reference to
perform our simulations.

\bibitem{repeat} B. Borstnik and D. Pumpernik, Europhys. Lett., \textbf{65},
290 (2004) and references therein.

\end{thebibliography}
\end{document}